\documentclass[twocolumn,
superscriptaddress,
prl,
showpacs,
preprintnumbers,
amsmath,amssymb]{revtex4}

\usepackage{graphicx,epsf,epsfig,ulem}% Include figure files
\usepackage{epsfig}
\usepackage{bm}% bold math
\usepackage{subfigure}
\usepackage{color}

\begin{document}

\title{Engineered valley-orbit splittings in quantum confined nanostructures in silicon}

\author{R. Rahman*}
\affiliation{Advanced Device Technologies, Sandia National Laboratories, Albuquerque, NM 87185, USA}

\author{J. Verduijn*}
\affiliation{Kavli Institute of Nanoscience, Delft University of Technology, Lorentzweg 1, 2628 CJ Delft, The Netherlands}
\affiliation{Centre for Quantum Computation and Communication Technology, School of Physics, University of New South Wales, Sydney, New South Wales 2052, Australia}

\author{N. Kharche*}
\affiliation{Computational Center for Nanotechnology Innovations, Department of Physics, Rensselaer Polytechnic Institute, Troy, NY 12180, USA}

\author{G. P. Lansbergen}
\affiliation{Kavli Institute of Nanoscience, Delft University of Technology, Lorentzweg 1, 2628 CJ Delft, The Netherlands}

\author{G. Klimeck}
\affiliation{Network for Computational Nanotechnology, Purdue University, West Lafayette, IN 47907, USA}
%\affiliation{Jet Propulsion Laboratory, California Institute of Technology, Pasadena, CA 91109, USA}

\author{L. C. L. Hollenberg}
\affiliation{Centre for Quantum Computation and Communication Technology, School of Physics, University of Melbourne, VIC 3010, Australia}

\author{S. Rogge}
\affiliation{Kavli Institute of Nanoscience, Delft University of Technology, Lorentzweg 1, 2628 CJ Delft, The Netherlands}
\affiliation{Centre for Quantum Computation and Communication Technology, School of Physics, University of New South Wales, Sydney, New South Wales 2052, Australia}

\date{\today}

\begin{abstract}
An important challenge in silicon quantum electronics in the few electron regime is the potentially small energy gap between the ground and excited orbital states in 3D quantum confined nanostructures due to the multiple valley degeneracies of the conduction band present in silicon. Understanding the ``valley-orbit'' (VO) gap is essential for silicon qubits, as a large VO gap prevents leakage of the qubit states into a higher dimensional Hilbert space. The VO gap varies considerably depending on quantum confinement, and can be engineered by external electric fields. In this work we investigate VO splitting experimentally and theoretically in a range of confinement regimes. We report  measurements of the VO splitting in silicon quantum dot and donor devices through excited state transport spectroscopy. These results are underpinned by large-scale atomistic tight-binding calculations involving over 1 million atoms to compute VO splittings as functions of electric fields, donor depths, and surface disorder. The results provide a comprehensive picture of the range of VO splittings that can be achieved through quantum engineering.   

\end{abstract}


\maketitle 

The ability to generate and manipulate three dimensionally confined quantum states in silicon down to the single electron regime is a much sought-after goal both in semiconductor quantum computing (QC) and quantum electronics. Silicon has not only been the primary platform of the semiconductor industry for over half a century, but in the quantum regime it also offers the advantage of long spin coherence times \cite{Tryshkin.prb.2003} necessary for QC. Over a decade, steady progress has been made towards realizing quantum dot (QD) \cite{Vincenzo.pra.1998, Friesen.prb.2003} and donor \cite{Kane.nature.1998, Vrijen.pra.2000, Hollenberg.prb.2004, Hill.prb.2005} based single electron states to encode and process quantum information. However, non-trivial challenges towards quantum electronics in Si arise from the existence of multiple conduction band (CB) valley degeneracies in Si, which multiples the orbital degrees of freedom. Since qubits require a two-level spin system well isolated in energy from other states, a critical goal is to establish control over valley-orbit (VO) splitting by means of external perturbations and engineered quantum confinement. To date, a comprehensive understanding of VO splitting in Si has been hindered by a lack of consistent experimental data, with VO splittings measured over several orders of magnitude from ueV to meV \cite{Wilde.prb.2005, Lai.prb.2006, Goswami.NaturePhysics.2007, Jiang.apl.2010, Takashina.prl.2006}. Although a number of investigations made inroads towards a theoretical understanding of this problem  \cite{Boykin.apl.2004, Friesen.apl.2006, Kharche.apl.2007, Srinivasan.apl.2008, Saraiva.prb.2009}, a unified theory is still absent primarily due to the lack of realistic models of the region between Si and an insulator. In this Letter, we present measurements of the VO splittings in Si metal-oxide-semiconductor (MOS) nanostructures under various quantum confinement regimes engineered by internal and external electric fields. Million atom tight-binding (TB) simulations of the various confinement regimes are performed to explore the range of VO splittings possible in these structures, and to provide a unified theoretical underpinning of the experimental data. With this combined experimental and theoretical approach over a range of confinement regimes we are able to not only explain the range of VO splittings observed here, but in addition understand how to control the VO splitting in a range of device configurations for future applications.  

In general VO splittings are expected to depend critically on the details of the confinement potential, interfacial disorder, barrier material, lattice miscuts, substrate orientation, strain, electric and magnetic fields \cite{Boykin.apl.2004, Friesen.apl.2006, Kharche.apl.2007, Srinivasan.apl.2008, Saraiva.prb.2009}. Once understood, this suggests the ability to engineer VO interaction externally, and hence to directly tune the momentum space properties of Si.

Hall-bar experiments in strained Si quantum well with SiGe barrier have reported VO splittings of the order of ${\mu}$eVs \cite{Wilde.prb.2005, Lai.prb.2006, Goswami.NaturePhysics.2007}.
%\cite{Wilde.prb.2005, Lai.prb.2006, Goswami.NaturePhysics.2007}, which are several orders of magnitude less than theoretically predicted values for flat quantum wells \cite{Boykin.apl.2004, Boykin.prb.2004.1}. This discrepancy has been attributed to interface roughness and miscuts, as the electronic wavefunction samples a large part of the interface laterally. Indeed, 
It was shown that a strong vertical magnetic field, which reduces the lateral extent of the wavefunction and hence the exposure to disorder, could be used to obtain a relatively larger VO of 1.5 meV in the SiGe system \cite{Goswami.NaturePhysics.2007}. %3D confined quantum dot states are also likely to show an enhanced valley splitting for the same reason. 
Relatively little data exits for VO splittings in Si MOS QDs near the few electron regime. In Ref \cite{Jiang.apl.2010}, a VO splitting of 0.76 meV was reported recently. %Other measurements are expected in the near future as other groups have recently reached the few electron regime in Si MOS QDs \cite{Lim.apl.2009, HRL.arXiv.2009, Madhu.arXiv.2010}. 
A past measurement of VO in a SIMOX device showed an unusually high value of 23 meV \cite{Takashina.prl.2006}, which is yet to be explained conclusively, although a recent work has given some plausible arguments for the cause \cite{Saraiva.arXiv.2010}.    

\begin{figure}[htbp]
%\center\epsfxsize=3.4in\epsfbox{Fig1_new_v2.eps}
\center\epsfxsize=3.4in\epsfbox{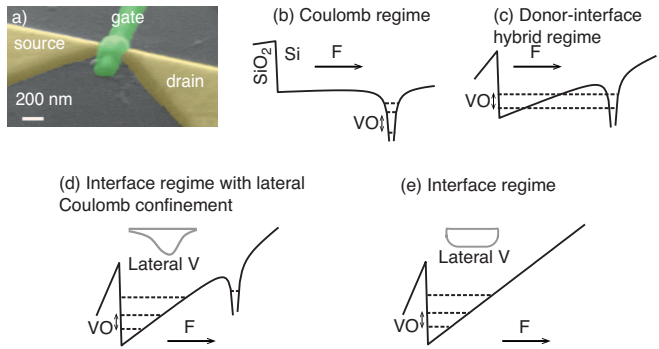}
\caption{a) SEM image of a FinFET device used in the experiments. b)-e) Schematic of the various confinement regimes showing the vertical confinement potential near the oxide interface. b) is a donor bound Coulomb confinement regime at low E-field. c) is a hybrid confinement regime between the donor and the interface well realized at higher E-fields when the two wells are lined up in energies and are strongly tunnel coupled. d) is an interfacial confinement regime realized at even higher fields, but laterally bound by the donor Coulomb potential. e) is a QD-like confinement regime realized at strong E-fields for device samples without any influence of the donor. The lateral confinement is provided by the residual barriers in the access regions. The insets show schematics of the lateral confinement potential in d) and e).  
}
\vspace{-0.5cm}
\label{fi1}
\end{figure}

Fig. 1(a) shows an scanning electron micrograph (SEM) image of a MOS FinFET device used in the experiments reported here. A silicon nanowire connected to source and drain leads forms the channel of the transistor. A second nanowire is deposited perpendicular to the channel as the gate electrode. A thin nitrided oxide layer separates the gate from the channel. %Current conduction takes effect through the corner effect \cite{Sellier.apl.2007}. 

This device has been used to realize and probe different confinement regimes in Si nanostructures. The measurements are based on excited state transport spectroscopy of a localized electron in a single quantum dot or donor. Most measurements utilize open system and investigate the quantum Hall effect in high B-field, and then extrapolate the measured value to the low B-field limit. Our method in essence provides an all-electrical means of measuring valley-orbit splitting.
%This device has been used to realize and probe different confinement regimes in Si nanostructures. The measurements are based on excited state transport spectroscopy through single quantum dot or donor states. Whereas most measurements of valley splitting use a high B-field, and then extrapolate the measured value to the low B-field limit, our method in essence provides an all-electrical means of measuring valley splitting. 
As noted, VO splittings change significantly with the applied electric field ranging from ${\mu}$eVs to tens of meVs. Hence, it is important to report the field at which VO measurements are done in experiments.
Some device samples contained a single Arsenic donor in the channel due to diffusion from the leads, and could be used to probe donor states \cite{Rogge.NaturePhysics.2008, Rogge.arXiv.2010}. Other samples without donors could be used to realize QD type states. % bperform excited state spectroscopy on single donor D0 \cite{Rogge.NaturePhysics.2008} and D- \cite{Rogge.arXiv.2010} states. To study quantum dot states in these devices, we have used a sample without any influence of dopant atoms in the channel. %All samples have a physical gate length of 60 nm. Since the gate defines the size of the channel, any confined state cannot be more extended than this length. %as the gate defines the size of the channel. %To study quantum dot states in these devices, we have used a sample with no donor atoms in the channel.

To provide a comprehensive theoretical basis for the experimental results we use a 10-band $sp^3d^5s^*$ nearest neighbor TB model involving over a million atoms, detailed in Refs \cite{Klimeck.cmes.2002, Klimeck.ted.2007, Rahman.prb.2009}. The method has been used successfully to accurately model a number of experiments on Si nanostructures \cite{Rahman.prl.2007, Rogge.NaturePhysics.2008, Rogge.arXiv.2010, Klimeck.ted.2007, Kharche.apl.2007}. %The full Hamiltonian with the electric field and closed boundary conditions is solved by a parallel Lanczos algorithm to capture the relevant number of quantum dot or donor states. 
A dangling bond passivated surface model has been used to represent the interfacial boundary \cite{Lee.prb.2004}, except when modeling the $\textrm{SiO}_2$ insulator layer explicitly, as shown in Fig. 4.

Fig. 1(b)-(e) show the schematic of the different confinement regimes investigated in this work. The gate potential generates a triangular well near the oxide interface. If a donor is present in the channel, an additional Coulomb potential well forms on top of the triangular well at some distance from the interface, making it possible to study different confinement regimes (Fig. 1 caption) in a number of device samples.%, as described in the figure caption. 
%Fig. 1(b) shows a schematic of the potential in the vertical (z) direction. The gate potential generates a triangular well near the oxide interface, where the quantum dot states are formed due to the lateral confinement from the residual barriers in the access regions between source/drain and channel. 

First, we discuss the three confinement regimes of Fig. 1 (b), (c) and (d) that are influenced by the presence of a donor. %Now we move onto the donor case where quantum confinement is more of a Coulomb-like potential. We have investigated how presence of single donors affect VO splittings in these structures. 
In the experimental device, the As donors are located less than 6 nm from the oxide interface and are subjected to fields of tens of MV/m. This donor-interface configuration has been proposed as an important system for implementing a donor-dot hybrid qubit %in which the donor bound electron can be selectively shuttled between the donor and the interface to take advantage of increased quantum control near the surface and increased spin coherence of the donor 
\cite{Smit.prb.2004, Martins.prb.2004, Calderon.prl.2006, Rahman.prb.2009}. 

\begin{figure}[htbp]
\center\epsfxsize=3.4in\epsfbox{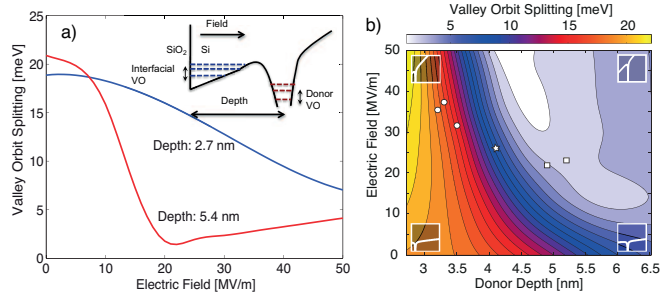}
\caption{%The states of the QD labeled by the lateral confinement quantum numbers and a two-valley index. The labels 'o' and 'e' represent 'even' and 'odd' valley combinations respectively as described in the text. 
a) Tight-binding calculations of VO splitting in Si in the presence of a single As donor as a function of field for two donor depths. The inset shows a 1D schematic of the confining potential with the donor. b) VO splitting as a function of donor depth and electric field obtained from TB. %The color code shows a range of VO splittings from sub meV to 22 meV depending on three confinement regimes. 
The white markers show the measured VO splitting for six device samples extracted from the measurements of Ref \cite{Rogge.NaturePhysics.2008}. }
\vspace{-0.5cm}
\label{fi2}
\end{figure}

For bulk donors at zero fields, all six valleys contribute to VO splittings. %The ground state is a symmetric superposition of Bloch functions from all six CB valley minima. 
Since the central cell potential varies from one donor species to another, so does the VO splitting \cite{Rahman.prb.2009}: in bulk the VO splitting of a bulk As donor is about 21 meV, compared to 12meV for a P donor. If an electric field is applied in the z direction, the weight of the wavefunction increases in the $k_z$ valleys and diminishes in the others. This wavefunction redistribution in momentum space causes a reduced VO splitting. Fig. 2(a) shows the calculated VO splitting as a function of the E-field for two different donor depths. %The inset shows a 1D schematic potential of the two-well system. 
At low E-fields, the VO splitting is about 20 meV, comparable to the VO splitting of a bulk As donor. %, but not exactly 21 meV due to the effect of the interface. 
As the E-field is increased, the donor states hybridize with interface states, and VO splitting reduces gradually. %to a value somewhere between that of a QD and a donor. 
At high enough E-fields, the electron is pulled to the interface, reducing VO splitting to a few meV, as expected of QD bound states at strong fields. Once the electron resides at the interface, the VO splitting varies linearly with the field, as shown by the red curve of a donor at 5.4 nm depth at fields above 30 MV/m. The blue curve is for a donor at a shallower depth of 2.7 nm. The change in VO is smoother because of stronger tunnel coupling between the two wells.

In Fig. 2(b), the VO splitting is plotted in color code as a function of donor depth and electric field. At low fields, the donor bound Coulomb-like regime of Fig. 1(b) is realized, whereas at high fields, the states are mostly interface bound, as in Fig. 1(d). %The red region of the plot shows a more Coulomb-like VO splitting, whereas the blue region shows a more dot-like splitting. 
%In between, there is the hybridized regime as labelled in Fig. 1(c). 
The measured VO splittings for six devices with donors at various depths and fields are mapped on this figure as white markers. The data points sample out all three confinement regimes. The measured VO data has been extracted from measurements of D0 excited state spectroscopy as reported in Ref \cite{Rogge.NaturePhysics.2008}, where the donor depths and applied fields were reported for various device samples. It is therefore possible to obtain a whole range of VO splitting in single donor devices ranging from 20 meV to an meV or even less. This means that VO splittings can be engineered through donor implantation depths and applied E-fields.     

\begin{figure}[htbp]
\center\epsfxsize=3.4in\epsfbox{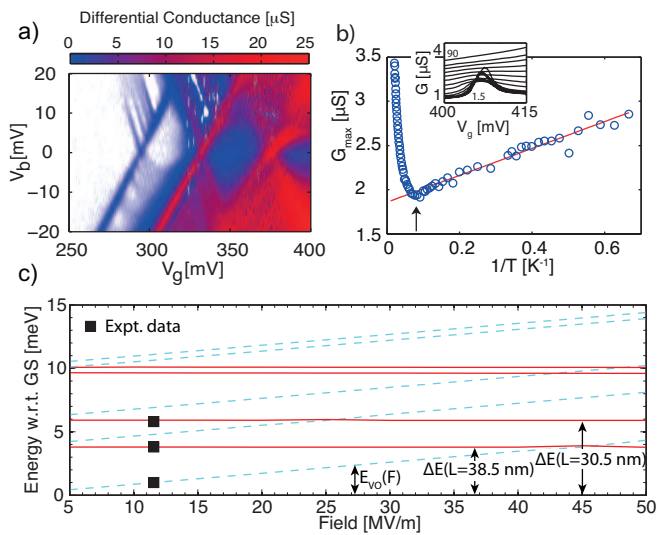}
\caption{a) Conductance vs. gate and bias voltage plot for a device without the influence of a donor (confinement regime Fig. 1(e)), showing blocked diamond region and tunneling through the QD states, characteristic of Coulomb blockade. b) Inset: Measured conductance data of the device as a function of gate voltage and temperature ($T$), between 1.5 and 90 K. The temperature steps are not all equal. Main plot: The peak maxima (extracted from the Inset) vs. $1/T$, showing a cross-over from single level transport (linear regime) to classical transport at higher $T$. c) Calculated spectrum of a MOS QD relative to the ground state, with lateral dimensions of 30.5 nm $\times$ 38.5 nm, as a function of the vertical electric field, showing VO split states, matched to the measured values. 
%c) Calculated spectrum of a MOS QD relative to the ground-state, with lateral dimensions of 30.5 nm \times 38.5 nm, as a function of the vertical electric field, showing VO split states and confinement states, matched to the measured values.}
} 
\vspace{-0.5cm}
\label{fi3}
\end{figure}

In Fig. 3, new measurements are shown for a device with a QD-like confinement regime described in Fig. 1(e). Unlike the previous cases, the lateral confinement here is provided by the residual barriers in the access regions. Hence, these are more extended interface bound states, as expected of MOS QDs. The measured experimental data for transport through the gate-field confined states is shown in Fig. 3(a). The low charging energy of about 10 meV, determined by the height of the Coulomb blockade diamonds, indicates there are no dopants present. Typical charging energies for dopant bound states range from 30 to 50 meVs \cite{Rogge.NaturePhysics.2008}. Furthermore, it was found that spin filling is consistent with the first diamond corresponding to the the first electron \cite{Sellier.prl.2006}.
%Fig. 1(d) shows an FFT of a ground state tight-binding wavefunction formed in the channel. The plot shows the two $k_z$ conduction band valleys contributing to this state, as expected in the presence of a vertical electric field. 
  
%In Fig. 2(c), we show the measured conductance data of the quantum dot states as a function of source-drain and gate biases. The excited states of the dot are clearly visible as conductance traces in the leftmost Coulomb diamond. From the data, we extracted the energy gaps of the excited states relative to the ground state. 

To make sure there is no unobserved low lying state between the ground and the first excited state in the stability diagram, we measured the temperature dependence of the low bias trace versus gate voltage. This method gives us the position of the lowest excited state \cite{Foxman.PRL.1994}. The inset of Fig. 3(b) shows the temperature dependence of the traces. In the temperature regime where the peak conductance is a linear function of the inverse temperature, $1/T$, there is transport through a single quantum state. Here the peak conductance increases with decreasing temperature (Fig. 3(b)). At higher temperatures, multilevel based classical transport mechanisms dominate, and at even higher temperatures, thermally activated transport dominates. The position of the cross-over point between the quantum regime and classical regime is determined by the position of the first excited state. Here, we found VO to be $\sim12$ K, or $\sim1 \pm 0.1$ meV. 
%Note that this value is smaller than the value expected for a gate-confined state using the (light) bulk effective mass of 0.2m$_e$, which would be $\sim0.5$ meV.

% ++++
% ++++

% This may not be the right place, but we have put a paragraph discussing the experimentally found QD VO values: It may be that the VO in the QD samples is somewhat smaller than we expect based on typical field values for the dopant samples and the TB VO coefficient. Is may be that this is due to disorder as for example the gate oxide may be "wobbly" and therefore there may exist local variations in the field (local, but not atomic!). We don't know what effect this has.

% ++++
A variety of factors such as orbital confinement, valley polarization, and VO interaction compete to determine the electronic structure of Si QDs. We have therefore performed TB simulations for this QD-like confinement regime as shown in Fig. 3(c). Since the effective lateral dimensions and the microscopic E-fields ($F$) in the channel are not known, we considered a range of values for these parameters, and obtained best agreement with the measured data %lateral confinement lengths and vertical E-fields. The lowest three energy gaps show best agreement with the measured data 
for lateral confinement lengths of $L_{x}=30.5$ nm, $L_{y}=38.5$ nm and $F=11.6$ MV/m. An instance of a simulated energy spectrum is shown in Fig. 3(c) as a function of the E-field. %for an asymmetric lateral confinement dimensions of 40 nm $\times$ 50 nm. 
The energy of the excited states are plotted relative to the ground state energy. The orbital states of the dot arising from the same valley configuration appear as flat lines with the vertical E-field. Their energies are primarily determined by the lateral dimensions of the box. The shorter the confinement lengths, the higher are the orbital energies, consistent with a 2D particle in a box model. The VO split states appear as tilted lines, showing a linear dependence on the E-field. An 1 meV VO splitting is obtained at $F=11.6$ MV/m independent of the lateral dimensions. The lowest three measured energy gaps are superimposed on this plot as white markers. To check that the 1 meV state is not an orbital state of the same valley configuration, we performed a back-of-the-envelope calculation to estimate the confinement length it would correspond to if it were indeed an orbital state, and obtained $L=77$ nm. Comparing with the channel and the gate lengths, the QD states are expected to have confinement lengths of less than 60 nm. Therefore, the 1 meV state is indeed a VO split state. %, and the TB simulations confirm this as well.

%due to a different valley configuration from the ground state appear as tilted lines with the E-field. The lowest tilted line is the VO splitting of interest. This means that the VO splitting increases smoothly with the field, as the dot electron is more sharply confined at the interface. %The lowest tilted line represents the VO splitting. %Extrapolation of this line to zero E-field yields a VO splitting of about 0.1 meV for this surface passivated model.

Since the VO split states can cross the orbital states of the dot at higher E-fields, the level arrangements in Si dots can vary from one experiment to another depending on the E-field and effective lateral confinement lengths. Hence, it is important to determine the E-field when reporting a VO measurement. The first excited state energy gap of the QD changes slope with E-field at this crossing point, which can be used as a reference point for measuring VO and orbital splittings in experiments in which the vertical field can be tuned.

\begin{figure}[htbp]
\center\epsfxsize=3.4in\epsfbox{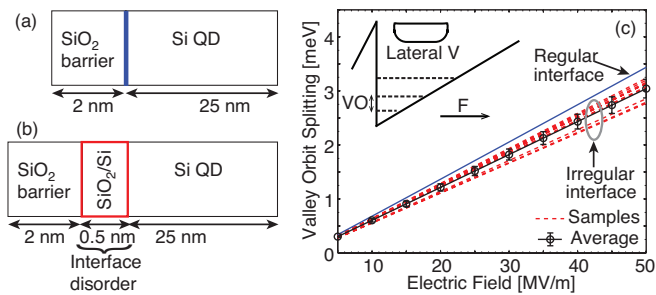}
\caption{%a) Tight binding calculations of VO splitting in a Si QD with an $\textrm{SiO}_2$ barrier as a function of the potential step at the conduction band. The vertical field values vary from 10 MV/m to 50 MV/m. The dashed horizontal lines are for a surface passivated model of the structure without the $\textrm{SiO}_2$ barrier. b) 
(c) TB simulations of VO splitting as a function of the vertical field for an ordered (a) and disordered (b) surface. %The error bars show the standard deviation in VO splittings over 10 randomly generated samples.
}
\vspace{-0.5cm}
\label{fi4}
\end{figure}

VO splittings in realistic systems are likely to be influenced by the atomistic details of the $\textrm{SiO}_2-\textrm{Si}$ interface. We performed TB calculations to
investigate the role of interface disorder on VO splittings in MOS QDs. %The TB method has proven to be a powerful tool to model disorder at the atomic scale \cite{Kharche.apl.2007,  Srinivasan.apl.2008}. 
We have used a virtual crystal (VC) 4 band $sp^3$ TB model of $\textrm{SiO}_2$ \cite{Luisier.sse.2010}. %The onsite energies and hopping parameters of this model have been optimized to fit some critical features of $\textrm{SiO}_2$ such as the bandgap and effective masses known from existing experimental measurements. 
Since the oxide is amorphous or highly disordered in reality, this is an approximation. However, given that ab-initio studies have shown some crystalline structure in the oxide near the interface \cite{Hybertsen.prl.2005} where the exponential tail of the QD wavefunctions reside, this model is expected to provide a good qualitative picture.

In Fig. 4(a), we included 2 nm width of a VC $\textrm{SiO}_2$ in the simulation domain along with Si. To simulate disorder, we have used an additional unit cell of $\textrm{SiO}_2$ at the the $\textrm{SiO}_2-\textrm{Si}$ interface (Fig. 4(b)), and replaced some of the virtual $\textrm{SiO}_2$ atoms with Si randomly. This in effect creates a local variation in $\Delta E_c$, and an overall decrease in the average $\Delta E_c$ at the interface. Simulations of 10 randomly generated samples show a slightly decreased VO splitting as a function of field. The standard deviation of VO splittings is represented by the error bars. Thus interface disorder does not have a significant effect in the calculated VO splittings in this model. In the field regimes investigated, the vertical E-field creates the most significant change in VO splittings.

We have investigated experimentally and theoretically the valley-orbit splitting in Si MOS devices under a range of quantum confinement conditions and provide a unified description of the large range of VO gaps observed in QDs and in hybrid single donor states. Large scale atomistic tight-binding simulations confirm quantitatively the range of observed VO splittings, and shed light on the role of interfaces, disorder, and electric fields. In Si MOS QD states, VO splittings are shown to increase with the vertical E-field, and can be influenced by atomistic disorder at low E-fields. 
Presence of single donors in these devices can yield VO splittings from 20 meVs to sub meVs depending on the field. %Since VO splittings vary significantly with applied field, it is essential to report the field value at which VO splittings are measured in experiments. 
This work enhances our understanding of VO interaction induced energy gaps in Si nanostructures, a critical consideration for Si qubits.    

\begin{acknowledgments}
Sandia is a multiprogram laboratory operated by Sandia Corporation,
a Lockheed Martin Corporation, for the United States Department of Energy under Contract No. DEAC04-
94AL85000. NEMO-3D was initially developed at JPL, Caltech under a contract
with NASA. NCN/nanohub.org computational resources were used. This research was conducted by the Australian Research Council Centre of Excellence for Quantum Computation and Communication Technology (project number CE110001029), NSA and ARO (contract number W911NF-08-1-0527).
%LCLH acknowledges the support of the Australian Research Council, NSA and ARO (contract number W911NF-08-1-0527). SR acknowledges support from the EC FP7 FET-proactive NanoICT projects MOLOC (215750) and AFSiD (214989). %, the Dutch Fundamenteel Onderzoek der Materie FOM. 
\end{acknowledgments}

Electronic address: rrahman@sandia.gov.
RR, JV and NK contributed equally to the research.

\end{document}